\documentclass[%
apl
 sd,%
 amsmath,amssymb,
 reprint,%double  column
]{revtex4-1}

\usepackage{graphicx}% Include figure files
\usepackage{dcolumn}% Align table columns on decimal point
\usepackage{bm}% bold math
\usepackage{color}
\definecolor{fgred}{rgb}{0.85 ,0 ,0}

\begin{document}

%\preprint{AIP/123-QED}

\title[Sample title]{High density NV sensing surface created via He$^{+}$ ion implantation of  $^{12}$C diamond }% Force line breaks with \\
%\thanks{Footnote to title of article.}

\author{Ed E. Kleinsasser$^{1}$}
\email{edklein@uw.edu}
\author{Matthew M. Stanfield$^{2}$}
\author{Jannel K.Q. Banks$^{2}$}
\author{Zhouyang Zhu$^{3,4}$}
\author{Wen-Di Li$^{3,4}$}
\author{Victor M. Acosta$^{5}$}
\author{Hideyuki Watanabe$^6$}
\author{Kohei M. Itoh$^{7}$}
% \altaffiliation[Also at ]{Department of Electrical Engineering, University of Washington, Seattle, Washington 98195-2500, USA}%Lines break automatically or can be forced with \\
\author{Kai-Mei C. Fu$^{1,2}$}%
\email{kaimeifu@uw.edu}
\affiliation{$^{1}$Department of Electrical Engineering, University of Washington, Seattle, Washington 98195-2500, USA  \\
$^{2}$Department of Physics, University of Washington, Seattle, Washington 98195-1560, USA \\
$^{3}$HKU-Shenzhen Institute of Research and Innovation (HKU-SIRI), Shenzhen, China \\
$^{4}$Department of Mechanical Engineering, The University of Hong Kong, Pokfulam, Hong Kong, China \\
$^{5}$Department of Physics and Astronomy, Univeristy of New Mexico, Center for High Technology Materials, Albuquerque, NM, 87106, USA \\
$^{6}$Diamond Research Laboratory, National Institute of Advanced Industrial Science and Technology (AIST), Tsukuba Central 2-13, 1-1-1, Umezono, Tsukuba, Ibaraki 305-8568, Japan\\
$^{7}$School of Fundamental Science and Technology, Keio University, 3-14-1 Hiyoshi, Kohoku-ku,
Yokohama 223-8522, Japan}
%Authors' institution and/or address%\\This line break forced with \textbackslash\textbackslash

%\affiliation{}

\date{\today}% It is always \today, today,

\begin{abstract}
We present a promising method for creating high-density ensembles of nitrogen-vacancy centers with narrow spin-resonances for high-sensitivity magnetic imaging.  Practically, narrow spin-resonance linewidths substantially reduce the optical and RF power requirements for ensemble-based sensing. The method combines isotope purified diamond growth, {\it in situ} nitrogen doping, and helium ion implantation to realize a 100 nm-thick sensing surface. The obtained $10^{17} \mathrm{cm}^{-3}$ nitrogen-vacancy density is only a factor of 10 less than the highest densities reported to date, with an observed spin resonance linewidth  over 10 times more narrow. The 200 kHz linewidth is most likely limited by dipolar broadening indicating even further reduction of the linewidth is desirable and possible.

%In this work we use He$^{+}$ ion microscope to implant a 100nm layer of N-doped, isotopically pure diamond layer grown via chemical vapor deposition. We then characterized the density and spin properties of the NV centers using confocal microscopy and optically detected magnetic resonance. The resulting NV densities are on the order of 10$^{17} \mathrm{cm}^{-3}$ which is only a factor of 10 less than the highest densities reported to data, but with superior spin properties. We found an effective dephasing time $T^*_2 = 1.5 \mu$s. Overall this appears to be a promising method for creating high-density ensembles of NV centers for high sensitivity magnetic sensing. 
%
%Valid PACS numbers may be entered using the \verb+\pacs{#1}+ command.
\end{abstract}

%Use showkeys class option if keyword
\maketitle
 
%\graphicspath{{Figure_data/}}

The nitrogen-vacancy (NV) center in diamond is a versatile room-temperature magnetic sensor which can operate in a wide variety of modalities, from nanometer-scale imaging with single centers~\cite{balasubramanian2008nanoscale,  maze2008nanoscale} to sub-picotesla sensitivities using ensembles~\cite{wolf2014subpicotesla}. Ensemble-based magnetic imaging, utilizing a two-dimensional array of NV centers~\cite{steinert2010high, maertz2010vector, pham2011magnetic}, combines relatively high spatial resolution with high magnetic sensitivity. These arrays are ideal for imaging applications ranging from detecting magnetically tagged biological specimens~\cite{gould2014room,glenn2015single} to fundamental studies of magnetic thin films~\cite{rondin2014magnetometry}.  A key challenge for array-based sensors is creating a high density of NV centers while still preserving the desirable NV spin properties. Here we report on a promising method which combines isotope purified diamond growth, {\it in situ} nitrogen doping and helium ion implantation. In the 100 nm-thick sensor layer, we realize an NV density of $10^{17}~\mathrm{cm}^{-3}$ with a 200~kHz magnetic resonance linewidth. This corresponds to a a  DC magnetic sensitivity ranging from 170 nT  (current experimental conditions) to 10~nT (optimized experimental conditions) for a $1~\mu \mathrm{m}^{2}$ pixel and 1 second measurement time. 

Magnetic sensing utilizing NV centers is based on optically-detected magnetic resonance (ODMR)~\cite{budker2007optical, acosta2009diamonds, manson2006nitrogen}.  In the ideal shot-noise limit, the DC magnetic sensitivity is given by~\cite{rondin2014magnetometry}
\begin{equation}
\delta B_{\mathrm{ideal}} \simeq \frac{h}{g\mu_B}\frac{1}{ C}\frac1{\sqrt{\eta}}\sqrt{\frac{\delta\nu/\pi}{n_{NV}Vt}}
\label{eq:sensitivity}
\end{equation}
%shot-noise sensitivity- will use this later
%\begin{equation}
%\delta B \approx \frac{h}{g\mu_B}\frac{1}{\sqrt{I_0VT n_{NV} }}\frac{\delta \nu}{C} = A\frac{\delta\nu}{\sqrt{N}}}
%\%label{eq:sensitivity}
%\end{equation}
in which $h/g\mu_B = 36~\mu\mathrm{T/MHz}$, $C$ is the resonance dip contrast, $\eta$ is the photon collection efficiency, $\delta \nu$ is the full-width at half maximum resonance linewidth, $n_{NV}$ is the density of NV centers in imaging pixel volume $V$,  and $t$ is the measurement time. From Eq.~\ref{eq:sensitivity}, it is apparent that to minimize $\delta B_{\mathrm{ideal}}$ for a given linewidth $\delta\nu$, one would like to maximize the NV density $n_{NV}$. Increasing $n_{NV}$, however, can also increase $\delta\nu$.  For example, lattice damage during the NV creation process can create inhomogeneous strain-fields~\cite{fang2013high}.  More fundamentally, eventually NV-NV and NV-N dipolar interactions will contribute to line broadening. This dipolar broadening,  $\delta\nu_{dp}$, is proportional to the nitrogen density $n_{N}$ \cite{wang2013spin, taylor2008high}. Since $n_{NV}$ is typically proportional to $n_{N}$, we can divide $\delta_\nu$ into two components, $\delta\nu = \delta\nu_0 + \delta\nu_{dp} =  \delta\nu_0  + A n_{NV}$, to obtain
\begin{equation}
\delta B_{\mathrm{ideal}} \simeq \frac{h}{g\mu_B}\frac{1}{C}\frac1{\sqrt{\eta V t\pi}}\sqrt{\frac{\delta\nu_0}{n_{NV}}+A}
\label{eq:sensitivity2}
\end{equation}
in which $\delta\nu_0$ depends on factors independent of NV density (e.g. hyperfine interaction with lattice nuclei, inhomogeneous strain fields). The second term $A$ is due to the dipolar contribution to the linewidth and will depend on the ratio of $n_N$ to $n_{NV}$.  Eq.~\ref{eq:sensitivity2} indicates there is not a single optimal NV density for  maximum sensitivity, but a minimum one, {\it i.e.} $n_{NV} \gtrsim \frac{\delta\nu_0}{A}$. 

However, there are practical reasons why magnetometry performance is higher for lower densities. By minimizing ODMR linewidth, we minimize both the excitation optical power (linear scaling with $\delta\nu$)  and RF power (quadratic scaling with $\delta\nu$)  requirements for the measurement.  Any technical noise proportional to the collection rate (e.g. laser intensity fluctuations) will also be improved with lower densities~\cite{kehayias2014microwave}. Finally, reduced densities resulting in reduced photon count rates will maximize the measurement duty cycle, minimizing detector dead time/readout time. Thus, a reasonable method to optimize $\delta B$ is to first minimize $\delta\nu_0$ and then increase $n_{NV}$ until the density independent and dipolar contributions to the sensitivity become comparable.

To minimize $\delta\nu_0$, this work utilizes nitrogen that is incorporated {\it in situ} during diamond growth on a (100)-oriented electronic grade substrate (Element Six, $n_{N, substrate} < 1$~ppb). {\it In situ} doping theoretically enables  uniform-in-depth nitrogen incorporation in the 100 nm thick sensor while avoiding lattice damage caused by (more standard) nitrogen ion implantation. Additionally, we utilize isotope purified $^{12}$C to eliminate $\delta\nu$ broadening due to the NV hyperfine coupling to $^{13}$C~\cite{ohashi2013negatively, itoh2014isotope}.  A linewidth $\delta\nu_0$ of tens of kHz is expected for these samples based on similar growth conditions for very low density nitrogen samples~\cite{ishikawa2012optical}.  A nitrogen density of 0.1-1 ppm was targeted during growth to obtain $\delta\nu_{dp}\approx\delta\nu_0$~\cite{taylor2008high, wang2013spin}.

Next the sample was implanted with $\mathrm{He}^{+}$ ions to create lattice vacancies. %It has been found that ensembles of NV centers create by implanting with lighter helium ions have better optical properties relative to those implanted with heavier nitrogen ions for similar implantation and annealing conditions~\cite{huang2013diamond}. 
He$^+$ implantation into a uniformly doped layer produces a uniform layer of NV centers with a controllable sensor thickness.
The method also provides independent handles on both nitrogen and vacancy densities to optimize NV formation. This is impossible with N$^+$ implantation alone where typically dozens of vacancies are created for every implanted N$^+$ ion. Different areas of the sample were implanted with ion doses ranging from 10$^{9}$-10$^{13}$ cm$^{-3}$ at acceleration voltages of 15, 25, and 35 keV. After implanting, the sample was annealed at 850 $^{\circ}$C for 1.5 hours in an Ar/H$_{2}$ forming gas to allow the vacancies to diffuse and bind with the doped nitrogen in the lattice forming NV centers. A second 24 hour anneal at 450 $^{\circ}$C in air was performed to convert NV centers from the neutral (NV$^{0}$) to the negative (NV$^{-}$) charge state \cite{fu2010conversion}.

%--------------------------------------------------------  FIGURE 1:  CARTOON OF DIAMOND WITH CONFOCAL SCAN OF 3 SQUARES -----------------------------------------------------------%
\begin{figure}[h]
\includegraphics{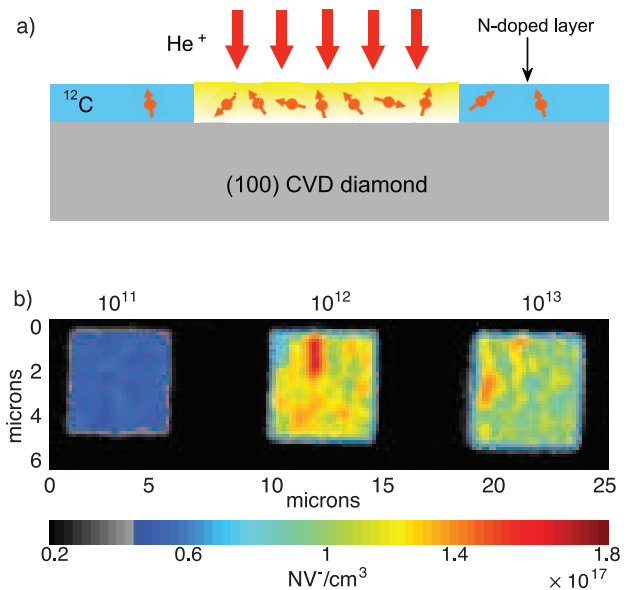}
\caption{\label{fig:cartoon} a) Schematic of the diamond sample illustrating the 100 nm $^{12}$C isotopically pure layer implanted with helium ions. b) Confocal scan of 5~$\mu\mathrm{m}^{2}$ implanted squares. Excitation at 532 nm with 1 mW power, collection band from 650-800 nm. From left to right, the squares were implanted with ion doses of $10^{11}, 10^{12}$, and $10^{13} ~\mathrm{cm}^{-2}$ at an acceleration voltage of 15 keV.} 
\end{figure}
%--------------------------------------------------------------------------------------------------------------------------------------------------------------------------------------------------------------------------------------------%

%--------------------------------------------------------   SECTION:  MEASURED NV DENSITY  -------------------------------------------------------------------------------------------------------------------------%

To characterize the NV$^{-}$ density, photoluminescence intensity from the $\mathrm{He}^{+}$ implanted squares were compared to single NV centers in a control sample. The 2D density was calculated with the known excitation spot size and converted to a 3D density utilizing the 100~nm sensor thickness. As only the negatively-charged state of the NV center is useful for magnetic sensing, room temperature PL spectra were used to confirm the synthesized centers were in the desired charge state.

Before ion implantation and annealing, the density of centers formed during growth ranged from 0.7-$3\times10^{15}$~cm$^{-3}$ (average value $1.5\times10^{15}~\mathrm{cm}^{-3}$). The range in density is due to uneven incorporation of nitrogen during CVD growth which will be discussed further below. Fig.~\ref{fig:cartoon} shows an NV$^{-}$ density map of three squares implanted with $10^{11}$, $10^{12}$, and $10^{13}$~$\mathrm{cm}^{-3}$ $\mathrm{He}^{+}$ ions at 15~keV after annealing. Experimentally we found that $n_{NV}$ for the three acceleration voltages varied by less than a factor of 2. This is consistent with SRIM calculations which show an average number of vacancies produced per ion of 30, 36, and 39, and an average ion range of 72, 112, and 135~nm,  for 15, 25, and 35 keV acceleration voltages, respectively.  All stopping ranges are within the 200~nm vacancy diffusion length~\cite{santori2009vertical} of the doped 100~nm layer. %The average number of vacancies/ion produced was 30, 35, and 39 for the 15, 25, and 35 keV respectively.%

During the implantation process, the entire sample was exposed to an unknown $\mathrm{He}^{+}$ radiation dose resulting in a background NV concentration of 0.1-1$\times 10^{16}$~cm$^{-3}$. Squares implanted with ion doses of $10^{9}$ and $10^{10}$ cm$^{-2}$ were indistinguishable from this background in most of the implanted areas. The optimal ion dose was 10$^{12}$ cm$^{-2}$ which resulted in an average $n_{NV}$ of 1$\times$10$^{17}$~cm$^{-3}$ corresponding to a 60-fold average increase over the unimplanted case. The obtained  density is only one order of magnitude lower than the highest densities reported~\cite{acosta2009diamonds, botsoa2011optimal}. These very high densities were obtained in high nitrogen doped ($>$100 ppm) diamond which exhibit significantly broader resonance lines (2~MHz)~\cite{acosta2009diamonds} due to the N-NV dipolar coupling. Densities of $10^{17} $~cm$^{-3}$ have also been obtained with N implantation and annealing~\cite{waxman2014diamond} which also exhibited several MHz linewidths.
 
\begin{figure}[h!]
\includegraphics{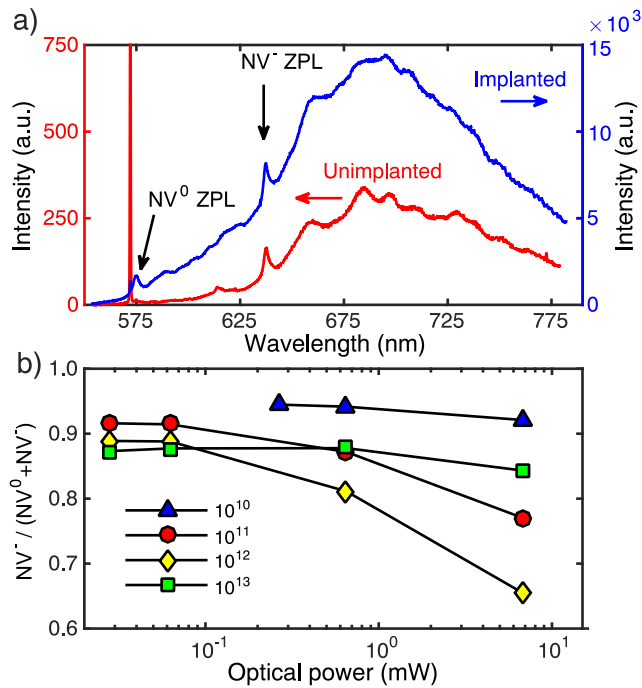} 
\caption{\label{fig:figure2} a) Spectra of unimplanted and implanted conditions illustrating the increase in photoluminescence after He$^{+}$ implanting (15~keV, $10^{12}~\textrm{cm}^{-3}$). Excitation at 532 nm with 1 mW power. b) Plot of the ratio of NV$^{-}$ to total NV vs. optical power. For the ratio, the difference in the relative weight of the NV$^{0}$ and NV$^{-}$ ZPL due to the different Huang-Rhys factors (approximately a factor of 2) has been taken into account \cite{davies1973luminescence, alkauskas2014first}. }
\end{figure}

Room-temperature spectra comparing the NV$^{-}$ zero-phonon-line photoluminescence intensities for the implanted and unimplanted cases show a similar increase ($\sim$40-fold) in NV density for the optimal implantation dose, as shown in Fig.~\ref{fig:figure2}a. Fig.~\ref{fig:figure2}b, shows the ratio of NV$^{-}$ to total NV (NV$^{-}$ + NV$^{0}$) for different optical powers. The high ratio at low intensities indicate the NV is predominately in the desired charge state in the absence of optical excitation. The decrease in ratio with increased power is consistent with photoionization effects reported previously \cite{manson2005photo}.
%--------------------------------------------------------------------------------------------------------------------------------------------------------------------------------------------------------------------------------------------%

%--------------------------------------------------------   FIGURE 2:  NV- 	vs  OPTICAL POWER BEFORE & AFTER IMPLANTING  , NV-/(NV- + NV0 -------------------------------------------------------%
%--------------------------------------------------------------------------------------------------------------------------------------------------------------------------------------------------------------------------------------------%

%--------------------------------------------------------   FIGURE 3:  LINEWITH, SENSITIVITY, OPTICAL CONTRAST, INTENSITY PLOTS  ------------------------------------------------------------------%
\begin{figure}[ht]
\includegraphics{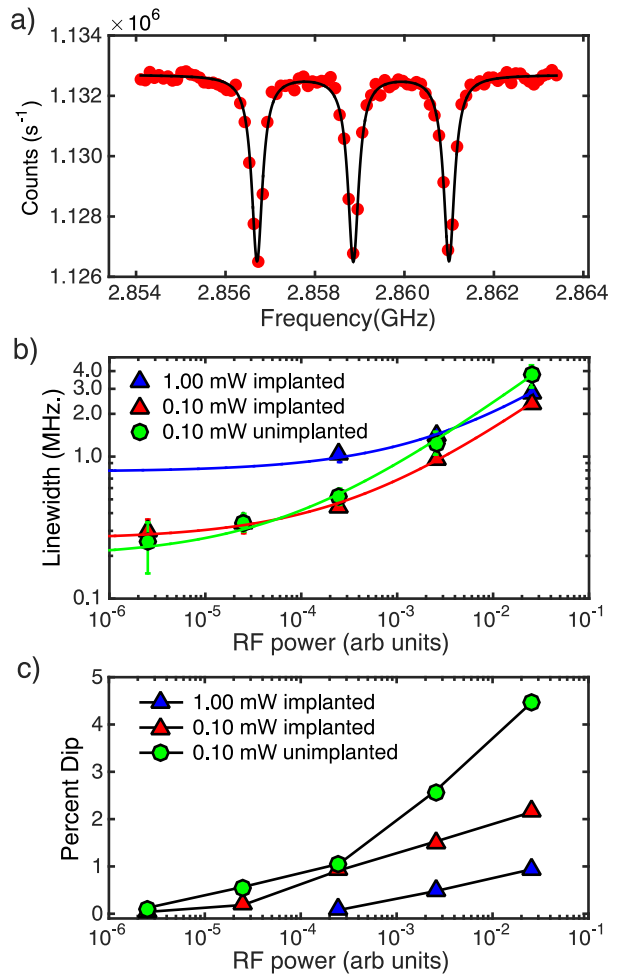} 
\caption{\label{fig:hyperfineLinewidth} a) ODMR scan of an implanted square with $\delta\nu = 290~$kHz. 30~$\mu$W of optical power and {-46}~dBm of RF power (measured at the output of our RF generator before the amplifier). b) Measured linewidth ($\delta\nu$ ) verses RF power. The error bars signify the 95 percent confidence interval of the width parameter for the Lorentzian fit function. c) Optical contrast (depth of the resonance dip) verses microwave power for unimplanted and implanted conditions. Implantation condition: 15~keV, 10$^{12}~\mathrm{cm}^{-2}$.}
\end{figure}
%--------------------------------------------------------------------------------------------------------------------------------------------------------------------------------------------------------------------------------------------%

%--------------------------------------------------------   SECTION:  CHARACTERIZING LINEWIDTH  --------------------------------------------------------------------------------------------------------------------%
Next we measured the ODMR linewidth, $\delta\nu$, of the  doped layer. Fig.~\ref{fig:hyperfineLinewidth}a shows an optically detected magnetic resonance (ODMR) spectrum for the $m_s=0 \leftrightarrow m_s= -1$  for one of the four NV crystal orientations. During the measurement, the NV centers are excited using a 532~nm continuous-wave laser while an RF field is swept through the electron spin resonance. Three dips are observed due to the hyperfine interaction of the NV electronic state with the $^{14}$N nucleus~\cite{acosta2009diamonds}. To determine $\delta\nu$, the ODMR fluorescence spectra were fit to the sum of three Lorentzian functions of equal amplitude and $\delta\nu$, with a fixed 2.17~MHz hyperfine splitting. 

Fig.~\ref{fig:hyperfineLinewidth}b shows a plot of $\delta\nu$ vs microwave power for the unimplanted and implanted conditions (15 keV, $10^{12}$~cm$^{-2}$). The data is fit to the theoretical model $ \delta \nu = \delta \nu_{RF = 0} + b\sqrt{P_{RF}}$ where $\delta\nu_{RF=0}$ is the intrinsic dephasing rate, $P_{RF}$ is the applied RF power, and $b$ a constant scaling to account for RF power broadening. The inhomogeneous spin relaxation time, $T_{2}^{*} = 1/(\pi \delta\nu_{RF=0})$, is determined from this fit. Experimentally we found that optical excitation powers below 100~$\mathrm{\mu}$W did not affect $T_{2}^{*}$. No measurable difference in $T_{2}^{*} $ was found between the unimplanted and 15 keV implantation cases, which both exhibit T$_2^*$ of 1.5~$\mathrm{\mu}$s ($\delta\nu \cong$ 200 kHz).  As little improvement in NV density was observed for higher implantation energies, detailed data on the effect of T$_2^*$ for implantation energies greater than 15 keV were not taken.

The observed 200~kHz linewidth for our dense ensemble of NV centers is 2-5 times more narrow than ensembles created via e-irradiation in natural diamond with 1 ppm N~\cite{acosta2009diamonds, acosta2013electromagnetically}, %\red{in which the dominant dephasing mechanism is attributed to hyperfine coupling with residual $^{13}$C.(need to be careful here)} 
suggesting that %even for the high NV density obtained in our sample, 
there is a significant benefit in utilizing isotope purified $^{12}$C.  The 200~kHz linewidth is, however, significantly broader than the 10~kHz inhomogeneous spread, due to the microscopic strain environment~\cite{rondin2014magnetometry}, between single NV centers grown in low nitrogen, $^{12}$C diamond~\cite{ishikawa2012optical}.  Given the minimal difference in T$_2^*$ for unimplanted vs implanted conditions, we attribute the dominant dephasing mechanism to dipolar  interactions between the NV centers and native nitrogen and/or between different NV centers.

Neglecting a small difference in g-factors between the NV and N~\cite{loubser1978electron}, we can obtain a rough estimate of the total density of paramagnetic impurities ($n_T = n_N + n_{NV}$) by looking at the characteristic magnetic dipole coupling between two centers, $\delta\nu \approx (\mu_0/4\pi \hbar) (g\mu_B)^2 n_T$~\cite{taylor2008high}. For a 200~kHz linewidth, we find an approximate density of 6$\times10^{17} \mathrm{cm}^{-3}$.  More detailed calculations taking into account NV interaction with the full N bath~\cite{wang2013spin} indicate the conversion efficiency may be much higher.  This numerical model found $1/T_2^* = 0.55 (\mathrm{\mu s}~\textrm{ppm}^{-1})f$ in which $f$ is the nitrogen density in ppm.  Using this expression, we find a nitrogen concentration of 2$\times10^{17} \mathrm{cm}^{-3}$. The average measured NV concentration for this implantation condition (15 keV, 10$^{12}~\mathrm{cm}^{-2}$) was 1$\times10^{17}$~cm$^{-3}$. The conversion efficiency of N$\rightarrow$ NV using He ion implantation thus ranges from 16 to 50\%, with the main uncertainty due to the accuracy of the dipolar model. 50\% is the maximum conversion for N$\rightarrow\mathrm{NV}^{-}$, assuming the extra electron needed to obtain the negatively charged NV state comes from substitutional nitrogen donors~\cite{collins2002fermi}.  Finally, we note that EPR measurements indicate 0.2 to 0.5\% of N incorporates as NV during -oriented CVD growth~\cite{edmonds2012production}, suggesting a 12-30\% conversion efficiency for our process. 

We now estimate the DC magnetic sensitivity of the engineered layer for a 1 second integration time. For a sensor biased at the steepest slope of the ODMR curve, the shot-noise magnetic sensitivity is given by~\cite{rondin2014magnetometry}

\begin{equation}
\delta B_{\mathrm{sn,cw}} \simeq \frac{h}{g\mu_B}\frac{1}{C}\frac{\delta\nu}{\sqrt{I_0t}}
\label{eq:dBsncW}
\end{equation}

in which $I_0$ is the detected photon count rate from the NV centers in the measurement pixel.
For the case of continuous-wave (CW) RF and optical fields, the realized sensitivity is a complex interplay between optical and RF power. The optical excitation power has counteracting effects on $\delta B_{\mathrm{sn, cw}}$, both increasing $I_0$ and $\delta\nu$ (Fig.~\ref{fig:hyperfineLinewidth}b). Similarly increasing the RF power will both increase $\delta\nu$ and $C$ (Fig.~\ref{fig:hyperfineLinewidth}c). For a 1~$\mathrm{\mu m}^2$ pixel, we find a  sensitivity of $\delta B_{\mathrm{sn,cw}} \simeq170$~nT at a 100 $\mathrm{\mu W}$ optical excitation intensity and an RF power corresponding to $C= 0.01$. This can be readily improved by a factor of $\sqrt 3$ utilizing a high-NA objective~\cite{jelezko2006single} and a further factor of 3 by driving all three hyperfine transitions simultaneously, resulting in a sensitivity of $\delta B_{\mathrm{sn, cw}}\simeq$  30~nT.

The sensitivity can be further improved utilizing pulsed techniques. In this case we can decouple the optical excitation from the spin manipulation, enabling the use of high optical powers for spin readout without adversely affecting the ODMR linewidth. In this scheme, the optimal spin-manipulation time is $\approx T_2^*$~\cite{taylor2008high}, resulting in a time-averaged photon count rate of $I_0 \rightarrow I_0 \tau_L/T_2^*$ in Eq.~\ref{eq:dBsncW}, in which $\tau_L$ is the optical read-out pulse length~\cite{rondin2014magnetometry}. The pulsed sensitivity, identical to Eq.~\ref{eq:sensitivity}, is given by

\begin{equation}
\delta B_{\mathrm{sn, pulsed}} \simeq \frac{h}{g\mu_B}\frac{1}{C}\frac{\sqrt{\delta\nu/\pi}}{\sqrt{I_0 \tau_L t}}.
\label{eq:dBsnpulsed}
\end{equation}

Using reasonable parameters ($\tau_L = 300$~ns~\cite{rondin2014magnetometry}, $C = 0.05$, $I_{0} = n_{NV}V\times10^5$~counts s$^{-1}$~\cite{jelezko2006single}) we estimate a sensitivity of $\delta B_{\mathrm{sn, pulsed}} \simeq 10~\mathrm{nT}$.

In future sensor fabrication, improvements to the magnetic sensitivity could be realized by utilizing (111) surfaces to obtain a single orientation of NV centers ($C \rightarrow 4C$)~\cite{michl2014perfect,lesik2014perfect, fukui2014perfect}. Additionally,  optimizing the initial nitrogen density $n_{N}$ such that  $\delta\nu_0 \approx \delta\nu_{dp} \approx 10$~kHz could result in a further $\sim$10-fold decrease in the ODMR linewidth. More critically, however, is the need to further improve the uniformity of N incorporation during CVD growth. In this work, initial nitrogen incorporation densities varied by a factor of 3-4. Theoretically, in a calibrated, stable imaging system, this deviation should not pose a problem. Practically, however, spatial variations over time (e.g. due to vibrations or thermal drift) will result in a false magnetic signal. It has been recognized that nitrogen incorporation during diamond growth is extremely sensitive to the growth plane \cite{samlenski1995inc, miyazaki2014atomistic} and thus surface steps on a 100 surface. By reducing the misorientation of the surface cut (typically 1\% in our samples), we expect to be able to enhance the incorporation homoneity.  High NV spatial uniformity combined with the realized optical and spin properties presented in this work is expected to result in a high-sensitivity magnetic imaging system for magnetically-tagged biological applications and the study of optical-scale magnetic phenomena.

{\it Acknowledgements} This work has been supported by a  University of Washington Molecular Engineering and Sciences Partnership grant. The work at Keio University has been supported by JSPS KAKENHI (S) No. 26220602 and Core-to-Core Program. VMA acknowledges support from NSF grant IIP-1549836. WDL was sponsored by NSF of China (Grant No. 61306123) and RGC of HKSAR (Grant No. 27205515). ZZ and WDL thank the facility support from Nanjing National Laboratory of Microstructures.

%\bibliography{fu_lab_bib,refs}% Produces the bibliography via BibTeX.
\bibliography{refs.bbl}% Produces the bibliography via BibTeX.

\end{document}